\documentclass[aps,pra,twocolumn,groupedaddress,floatfix]{revtex4}
\usepackage{bm} \usepackage{graphicx}
\usepackage{epsfig}
\usepackage{subeqn}

\begin{document}


\title{Dynamical pattern formation during growth  of a dual-species Bose-Einstein condensate}

\author{Shai Ronen} 
\affiliation{JILA and Department of Physics, University of Colorado, Boulder, CO 80309-0440}
\email{sronen@colorado.edu}
\author{John L. Bohn} 
\affiliation{JILA,  and Department of Physics, University of Colorado,
 Boulder, CO 80309-0440 }
\email{bohn@murphy.colorado.edu}

\author{Laura Elisa Halmo}
\affiliation{Department of Physics, Georgia Southern University, Statesboro,
GA 30460-8031}

\author{Mark Edwards}
\affiliation{Department of Physics, Georgia Southern University, Statesboro,
GA 30460-8031}

\affiliation{}


\date{\today}

\begin{abstract}

We simulate the growth of a dual species Bose-Einstein condensate
using a Gross-Pitaevskii equation with an additional gain term giving
rise to the growth. Such growth occurs during simultaneous
evaporative cooling of a mixture of two gases.  The ground state of a
dual condensate is normally either a miscible mixture, or an
immiscible phase with two spatially separated components. In a cigar
trap the ground state typically consists of one component in the center, and
the other component flanking it. Our simulations show that when the
condensates are formed in a cigar trap and the mixture is phase
separated, then the final state upon the end of the growth is
generally far from the true ground state of the system. Instead it
consists of multiple, interleaved bubbles of the two species. Such a
pattern was observed recently in an experiment by Wieman's group at
JILA \cite{Papp08}, and our simulations are in good qualitative
agreement with the experiment.  We explain the pattern formation as
due to the onset of modulation instability during growth, and study
the dependence of the final state pattern on various parameters of the
system.

\end{abstract}


\maketitle

\section{INTRODUCTION \label{intro}}

Like common fluid mixtures, dual species Bose-Einstein condensates
(BECs) may exhibit both miscible and immiscible phases. In the
miscible phase, the two species form a homogeneous solution, while in
the immiscible phase, they separate spatially \cite{Myatt97,
Esry97a}. In a BEC, the occurrence of these phases is controlled by
the ratio of the inter-component interaction to that of the
intra-component interactions.

The ground state structure of the immiscible phase is typically that
of a ball and shell where one species forms a shell around the other,
or that of two condensates lying side by side. In a quasi-1D trap, the
structure can be of one component flanked by the other. The exact
structure depends on the shape of the trap, the size of each
component, and the interactions between them. In general, the two
components tend to minimize the surface area that
separates them, as this surface contributes surface tension energy
\cite{Esry97a,Trippenbach00,Jezek02,Riboli02,Schaeybroeck08}.

By starting from the ground state of a miscible phase and increasing
the inter-component repulsion (relative to the intra-component
interactions), it is possible to make a transition from the miscible
to the immiscible phase. Such a change in the interactions can bring
about a modulation instability where the two components separate
locally and form a pattern of interleaved bubbles. Alternatively, such a
pattern emerges when half the population of one pure component is
suddenly transfered to another one, and the two components separate
\cite{Miesner99,Kasamatsu04,Kasamatsu06}.

A recent experiment \cite{Papp08} on a mixture of $^{85}$Rb and
$^{87}$Rb in an elongated cigar trap, observed an intriguing modulated
pattern upon the end of evaporative cooling (shown in figure 4 of that
reference).  The pattern is that of separated and interleaved
``bubbles'' of the two components, with up to 3-4 bubbles from each
component. This behavior is seen when the inter-component repulsion is
strong compared to the intra-component interactions, i.e, when an
immiscible phase is expected.. These specific patterns were observed
even under conditions when the interactions strengths were kept
constant during and after the evaporative cooling. Nor is there population transfer
involved, as the two components are different isotopes. Following
evaporative cooling one may then expect to obtain a binary condensate in
its ground state. But the observed modulated pattern does not look
like the ground state of the system.

In this paper we suggest that the emergence of this pattern has to do
with the dynamics of the formation of the binary condensate during the
evaporative cooling. When the condensates are formed they are
initially very small. Their mutual interaction (proportional to the
densities) is then weak and we expect them to coexist spatially, even
when the predicted phase for a Thomas-Fermi (large number of atoms)
regime is immiscible. As the condensates grow, spatial separation
is triggered by the growing inter-component repulsion. One may
expect this to bring about a modulation instability giving rise to the
final observed spatial pattern.

Our aim in this paper is to demonstrate this mechanism, predict the
conditions for the formation of the spatial structure, and determine
the factors that influence the number of bubbles formed. The paper is
organized as follows: in section~(\ref{sec:model}) we give the theoretical
background: we present the Gross-Pitaevskii gain model for a double
condensate, and review the theory of modulation instability mechanism. In section~(\ref{sec:numerical})
we present and discuss the results of our numerical simulations. We summarize our conclusions in section~(\ref{sec:conclusions}).

\section{ BINARY CONDENSATE GROWTH MODELING \label{sec:model}}

\subsection{Gross-Pitaevskii Gain model}

We are interested in the dynamics of the formation and growth of the
binary condensate during the evaporative cooling, in order to
understand the final spatial pattern that emerges. The description of
BEC growth is a challenging theoretical problem. In principle it is
possible to perform first-principles numerical simulations of the
relevant equations \cite{Drummond99, Blakie05}. However, a great deal
of physical insight can be obtained from a simpler model. In the first
instance, the thermal cloud is much more dilute than the
condensate. Therefore, it is plausible to treat the system as a two
fluid model, where the thermal component is assumed to be an ideal gas
and its interactions with the condensate are neglected. If one is only
interested in the condensate, then the role of the thermal component
is simply that of a particle reservoir feeding the growing
condensate. One thus arrives at the Gross-Pitaevskii gain model
\cite{Drummond00}. This is a Gross-Pitaevskii equation modified by a
linear gain term. For two components, the Gross-Pitaevskii gain
equations (GPGE) read:

\begin{subequations}
\label{GPGE}
\begin{eqnarray} 
\lefteqn{i \hbar \frac{\partial \Psi_1(\bm{r},t)}{\partial t}  =}  \\ \nonumber
& &  \left(- \frac{
\hbar^2 \nabla^2}{2 m_1}+ V_{ext}^{(1)}+g_1
|\Psi_1|^2+g_{12}|\Psi_2|^2+i \hbar \Gamma_1 \right )\Psi_1,
\end{eqnarray}

\begin{eqnarray}
\lefteqn{i \hbar \frac{\partial \Psi_2(\bm{r},t)}{\partial t}  = } \\ \nonumber 
& & \left(- \frac{ \hbar^2 \nabla^2}{2 m_2} +V_{ext}^{(2)}+g_2 |\Psi_2|^2+g_{12}|\Psi_1|^2
+ i \hbar \Gamma_2 \right) \Psi_2.
\end{eqnarray}
\end{subequations}
Here, $\Psi_1$ and $\Psi_2$ are the wavefunctions
corresponding to the two condensates.  The normalization of each
wavefunction is taken independently as
$\int d\bm{r}|\Psi_i(\bm{r})|^2=N_i$. $m_1$ and $m_2$ are the corresponding
atomic masses.  The condensates are assumed to be trapped in
axis-symmetric harmonic potentials,

\begin{eqnarray}
 V_{ext}^{(i)}(r,z)=\frac{1}{2}m_i \left[\omega_r^2 r^2+
 \omega_{z(i)}^2 (z-Z_i)^2 \right], \;\; i=1,2,
\end{eqnarray}

where $\omega_r, \omega_{z(i)}$ are the radial and transverse trapping
frequencies, and $Z_i$ the axial center of the traps.  We allow for
the possibility of different axial trapping frequencies for the two
condensates, as well as for a relative axial shift $Z_2-Z_1$ between
the centers of the traps, as occurs in the experiment \cite{Papp08}.

The intra-component coupling constants $g_i=4\pi\hbar^2 a_i/m_i$ are
characterized by the scattering lengths $a_1$ and $a_2$ between atoms
of the same species, while the inter-component coupling,
$g_{12}=2\pi\hbar^2a_{12}/m_{12}$, (with
$\frac{1}{m_{12}}=\frac{1}{m_1}+\frac{1}{m_2}$), is determined by the
scattering length $a_{12}$, where an atom of component 1
scatters from an atom of component 2.  The $\Gamma_i$'s are
the gain terms. They contribute to the growth of component $i$ within
timescale 1/$\Gamma_i$. Intuitively, the growth terms arise from boson
statistics, where the probability to add an additional atom to
condensate $i$ is proportional to the number of atoms already
condensed. In practice these are  phenomenological parameters and may be estimated by fitting to an experiment.

\subsection{ Modulation Instability Mechanism}

Following the analysis of Refs. \cite{Riboli02,Jezek02,Papp08}
we define the parameter
\begin{eqnarray}
\Delta=g_1 g_2/ g_{12}^2-1
\label{eq:phase}
\end{eqnarray}
that depends upon the ratio of the single species and interspecies
interactions. When the number of atoms is large enough the system may
be treated by the Thomas-Fermi approximation, which neglects the kinetic
energy.  Assuming $g_{12}>0$, there are two regimes: $\Delta>0$,
where the two condensates are miscible, and $\Delta<0$, where they are
immiscible due to the interspecies repulsion.  

It is important to bear in mind that this classification is correct
only when the Thomas-Fermi approximation is valid. It does not hold  for
small, trapped condensates, where the kinetic energy terms are
significant. In the numerical simulations presented in the next
section we shall show that in this case the condensates can initially
co-exist spatially even when $\Delta<0$. As they grow, spatial
separation sets in. To have a better understanding of the dynamics of
spatial separation, it is useful to consider an idealized system, of
two homogeneous condensates in a quasi one dimensional geometry
(relevant to the experiment \cite{Papp08}), with respective densities
$n_1$ and $n_2$, co-existing spatially at time $t=0$,
as discussed in detail in Refs.~\cite{Kasamatsu04,Kasamatsu06}. If $\Delta>0$, the system is stable. If
$\Delta<0$, the initial state is unstable, and an initially
homogeneous system will evolve dynamically. In this so called
modulation instability mechanism, the two components separate, and the
separation occurs in a wave pattern with a typical length scale which
depends on the relative interactions and densities. Regions of
over-density of one component appear in regions of under-density
of the other component, and vice versa. Nonlinearity accelerates the process and
brings about complete spatial separation. A pattern of interleaved
bubbles of the two components emerges. The final spatial pattern can
be very stable, since there is no room for movement to ``the sides'',
and the energy barrier for tunneling between the separated bubbles is
very big.

The analysis of the modulation instability in a dual component
BEC in  a quasi 1D  geometry was
done in Ref. \cite{Kasamatsu06}. There, the assumption of equal masses
$m_1=m_2 \equiv m$ has also been made, which considerably simplifies the
analysis.  The behavior of the system at small $t>0$ can be deduced
from the excitation spectrum. The eigenmodes of the system are
characterized by a longitudinal wave number $k$ and frequency
$\Omega$. For each eigenmode, the wave number $k$ is shared by the two
components
\footnote{ In \cite{Kasamatsu06} the authors allowed for two
independent wave numbers $k_1$ and $k_2$ for the two components. We
note that actually $k_1=k_2$ must hold for the solution to be an eigenmode. Their results are correct as long as one
takes $k_1$=$k_2$.}.  The instability manifests itself in a continuous
range of modes which have imaginary frequencies. The most unstable
mode is that with imaginary frequency of maximum absolute value. This
is the mode that will grow fastest. Its wavenumber $k_{max}$ is found
to be:
\begin{eqnarray}
\lefteqn{\hbar k_{max}=} \\ \nonumber
& & \frac{\sqrt{2}}{b} \left[ \sqrt{( g_2 n_2- g_1
n_1)^2-g_{12}^2 n_1 n_2}- g_2 n_2 - g_1 n_1 \right]^{1/2}
\label{kmax}
\end{eqnarray}

where $b$ is the radial harmonic oscillator length
$b=\sqrt{\frac{\hbar}{m \omega_r}}$, and we assume here $m_1=m_2=m$.

The imaginary frequency of this mode is found to be:

\begin{eqnarray}
G_{max}\equiv\Im(\Omega_{max})=\frac{\hbar k_{max}^2}{2 m},
\label{Gmax}
\end{eqnarray}
and the time scale for growth  is $T_{max}=2 \pi/G_{max}$.

\section{NUMERICAL SIMULATIONS \label{sec:numerical}}

\subsection{Mechanism of dynamical phase separation}

We have performed numerical simulations of the growth of the dual
$^{85}$Rb/$^{87}$Rb condensate based on Eq.~(\ref{GPGE}). In
correspondence with the experiment \cite{Papp08}, all our simulations
have the following fixed parameters. $a_{87}=99$ bohr is the
scattering length of $^{87}$Rb. $a_{85-87}=214$ bohr is the
inter-species scattering length. The trap parameters are: radial
frequency $2\pi \times 130$ Hz for both components; axial frequency $2
\pi \times 2.9$ Hz for $^{85}$Rb and $2 \pi \times 2.6$ Hz for
$^{87}$Rb.

\begin{figure}
\resizebox{3.5in}{!}{\includegraphics{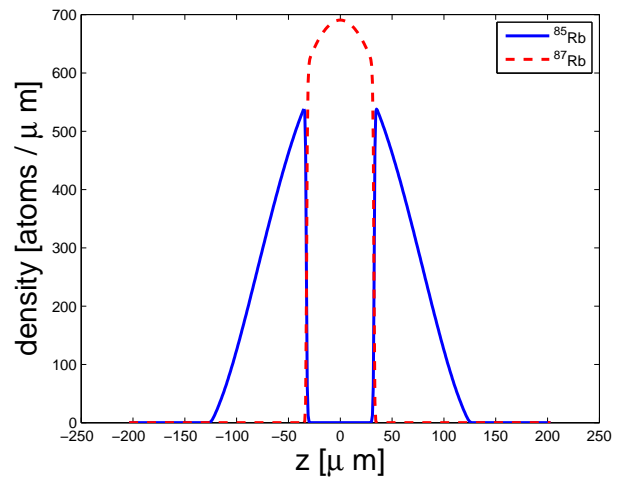}}
\caption {Calculated ground state of a dual ($^{85}$Rb/ $^{87}$Rb)
condensate in an immiscible phase, in a cigar shaped trap.  Shown are
the axial density profiles. The interaction parameters are:
$a_{85}=200$ bohr, $a_{87}=99$ bohr, $a_{85-87}=214$ bohr. The trap
parameters are: radial frequency $2\pi \times 130$ Hz for both
components; axial frequency $2 \pi \times 2.9$ Hz for $^{85}$Rb and $2
\pi \times 2.6$ Hz for $^{87}$Rb. The number of atoms is $5 \times
10^4$ in each component.
\label{fig:ground} }
\end{figure}

We first demonstrate that the true ground state of the system in an
immiscible phase is indeed quite simple, as described above. We
computed the ground state with $a_{85}=200$, and the other scattering
lengths fixed as above, so that $\Delta<0$. The ground state for this
system with 50,000 atoms in each component is shown in
Fig.~(\ref{fig:ground}). The calculation is 3D, and the figure shows
the density profile $\rho(z)$ along the long z-axis, where
$\rho(z)=\int \rho(x,y,z) \, dx \, dy$. We have also performed
additional computations which modeled the effects of a gravitational
sag that breaks the cylindrical symmetry and slightly shifts the centers of
masses of the the two components. In all cases, the ground state was
still similar in shape to that of Fig.~(\ref{fig:ground}).

\begin{figure*}
\resizebox{7.0in}{!}{\includegraphics{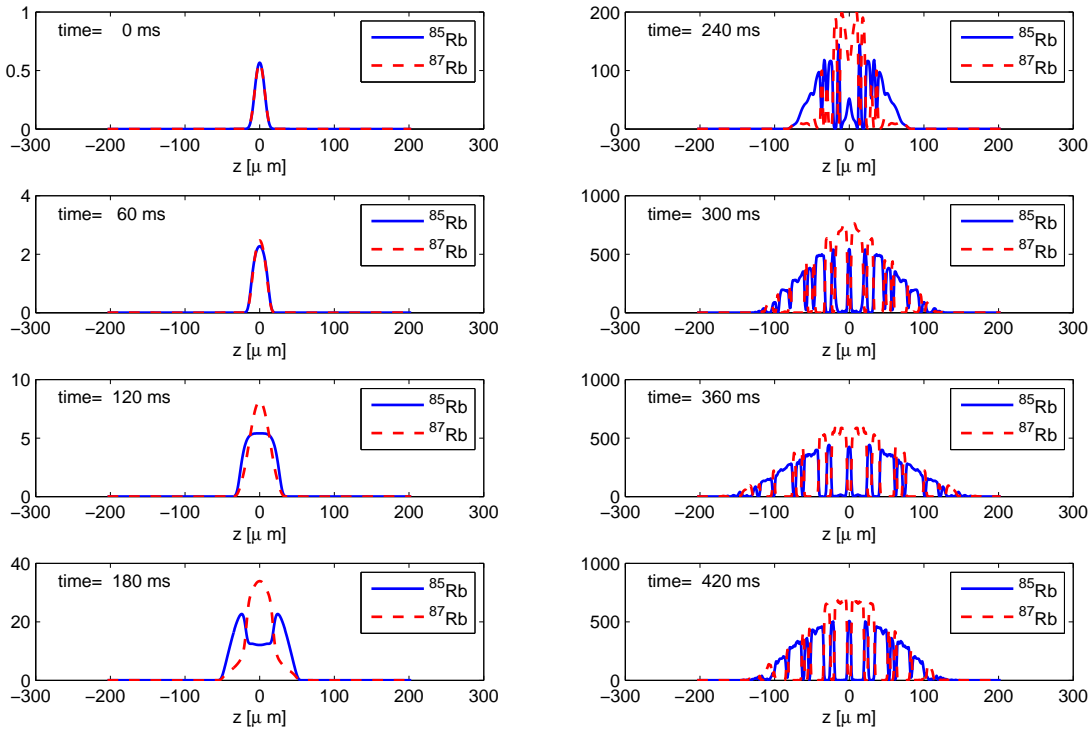}}
\caption {Dynamical growth of a dual BEC with the same interaction and
trap parameters as in Fig.~(\ref{fig:ground}). The initial state is
the ground state with 10 atoms in each component.
\label{fig:dyn3} }
\end{figure*}

We now investigate the consequences of a dynamical growth of a dual
condensate during evaporative cooling. The exact growth time of the
condensates in the experiment \cite{Papp08} is not known, but
has been roughly estimated to be between 100-500 ms
\footnote{J. Pino, private communication}. In our simulations we begin
with the ground state consisting of 10 atoms of each component, and
let the condensates grow for 300ms. We choose gain terms
$\Gamma_1=\Gamma_2=14.2 \text{sec}^{-1}$, giving rise to a final
population of 51,000 atoms in each component. The number of atoms
$N_i(t)$ of component $i$ at time $t$ can be calculated from:
$N_i(t)=N_i(0) e^{2 \Gamma_i t}$. We assume that after 300ms full
condensation has been achieved and the growth is terminated. The
$\Gamma_i$'s are then set to zero and further dynamics are atom number
conserving. Fig.~(\ref{fig:dyn3}) shows stages in the growth of the
dual condensate under these conditions. Animations of this evolution
are available at http://grizzly.colorado.edu/~bohn/movies/bubbles.htm

\begin{figure}
\resizebox{3.5in}{!}{\includegraphics{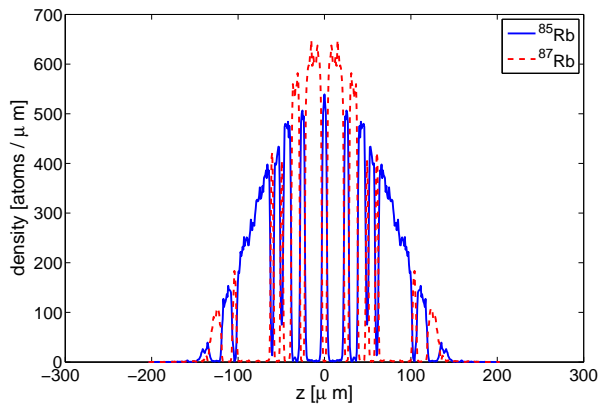}}
\caption {The final spatial pattern resulting from the dynamical
growth of the dual condensate as simulated in Fig.~(\ref{fig:dyn3}).
\label{fig:pattern3} }
\end{figure}

In the initial state (panel $t=0$), the number of atoms is very small
so that the mean field interactions are weak. Each condensate's shape
is essentially Gaussian and they are almost perfectly overlapping. As
the density grows (due to the imaginary gain term), the interactions
kick in (around 130ms, compare panels $t=120$ and $t=180$), leading to
phase separation at the center, where density is largest. After the
end of the growth (at 300ms, panel $t=300$) the modulation pattern is
observed to be quite stable, apart from a breathing motion, with the
components almost fully separated and forming multiple interleaved
bubbles, as shown in Fig.~(\ref{fig:pattern3}). Note the increasing
vertical scales, which account for growing particle numbers.

\begin{figure}
\resizebox{3.5in}{!}{\includegraphics{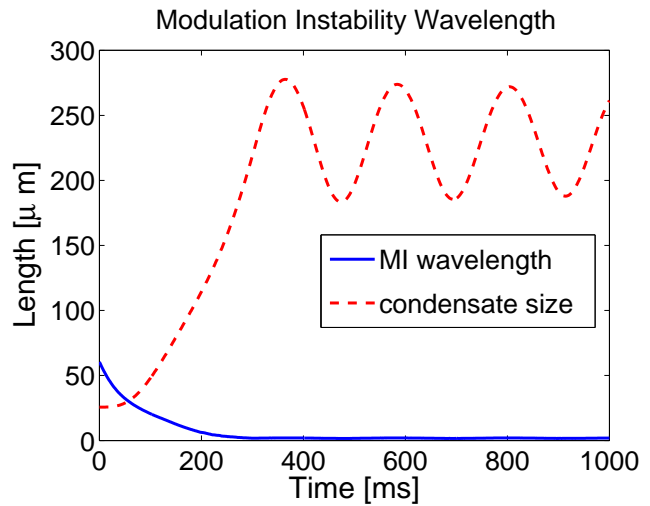}}
\caption {The modulation instability wavelength (solid line)
corresponding to the simulation of Fig~.(\ref{fig:dyn3}), as computed
from Eq.~(\ref{kmax}). This is compared with the r.m.s size of the total
density of both condensates along the trap long axis (dashed line). 
\label{fig:MI} }
\end{figure}

The onset of the immiscibility is triggered by the emergence of the
modulation instability.  The natural wavelength of the modulation
instability corresponds to the unstable mode with maximum imaginary
amplitude. This is $2\pi/k_{max}$ where $k_{max}$ is given by
Eq.~(\ref{kmax}); for the purpose of using this equation, which
assumes species with equal masses, we use the average mass of the two
isotopes of Rb, neglecting their small mass difference. Also, for our
non-homogeneous system, we use in it the maximum density of each
component. As the density of the condensates grows, their spatial
extent increases and at the same time the modulation instability
wavelength decreases. The decrease of the modulation instability
wavelength with time is shown in Fig.~(\ref{fig:MI}). Around ~60ms
after the beginning of the growth, the modulation instability
wavelength becomes comparable to the size of the condensate as
measured by the r.m.s of the density profile of the combined
condensates. The onset of dynamical separation is seen somewhat later
at ~130ms. This lag in time may be due to two factors. Firstly, the MI
wavelength Eq.~(\ref{kmax}) assumes a homogeneous condensate. In other
words, it applies when the local density approximation (LDA) is valid. On the other hand, the number
of atoms in each component at $t=130$ms is only about 400.  Thus, the
LDA approximation is not yet reliable. Secondly, as we discuss below,
there is a time scale associated with the growth of the unstable mode,
and this leads to a lag between the initial seeding of the instability
and its growth to an observable size. Nevertheless, the crossing of
the curves of the r.m.s size of the condensate and the modulation
instability wavelength indicates, to within a factor of two in time,
the onset of the dynamical instability.

As condensation proceeds and the density continues to grow, the
spatial extent of the condensates increases and at the same time the
modulation instability wavelength decreases further. This leads to the
continuous separation of the two components into smaller and smaller
``bubbles'', as can be seen in Fig.~(\ref{fig:dyn3}). The final size of
individual bubbles is determined by the point at which essentially complete
spatial separation between the two components is achieved, so that no
more sub-divisions are possible. The exact point at which this happens
depends strongly on the non-linear dynamics and is thus difficult to
estimate solely from the linear small perturbations analysis which is
behind Eq.~(\ref{kmax}). Typically, the final modulation instability
wavelength is much smaller than the size of the observed bubbles, but
this has no effect since at this point the bubbles are already almost
purely separated components.

\begin{figure}
\resizebox{3.5in}{!}{\includegraphics{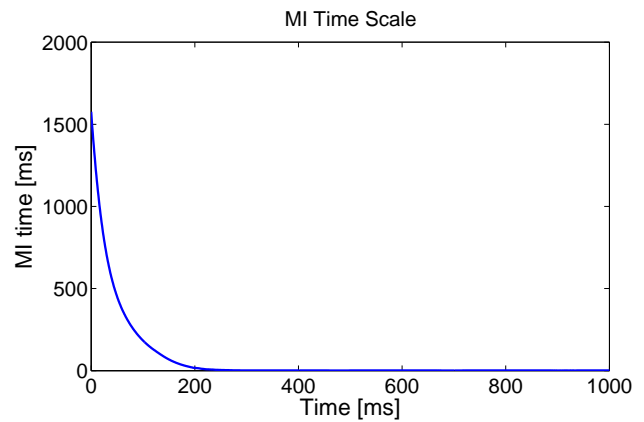}}
\caption {The timescale for growth of modulation instability
corresponding to the simulation of Fig.~(\ref{fig:dyn3}), as computed
from Eq.~(\ref{Gmax}). The horizontal axis is the simulation time, while
the vertical axis is the time scale for the growth of the most
unstable mode. This timescale is $T_{max}=2 \pi /G_{max}$.
\label{fig:MItime} }
\end{figure}

Regarding the temporal evolution, we can compare two time scales. One is the time
scale over which the condensate's population grows, and the other is the time scale
for the growth of the most unstable mode. This modulation instability
time scale depends on the density and thus is itself
time-dependent. Fig.~(\ref{fig:MItime}) shows the evolution of the
modulation instability time scale $T_{max}=\frac{2 \pi}{G_{max}}$ as
function of the simulation time, where $G_{max}$ is given by
Eq.~(\ref{Gmax}). As the density of the condensate grows, the
timescale for the growth of the most unstable mode is decreasing
exponentially.  Around ~130ms the time scale of growth of the unstable
mode becomes comparable to the time scale of growth of the condensates
themselves. As we have seen, this coincides with the point where
dynamical phase separation begins to be observable in the
simulation. With increasing simulation time, the modulation
instability growth time scale becomes very short, leading to the rapid
appearance of bubbles between 180ms and 240ms in
Fig.~(\ref{fig:dyn3}).

\subsection{Factors affecting the spatial pattern formation}

\begin{figure}
\resizebox{3.5in}{!}{\includegraphics{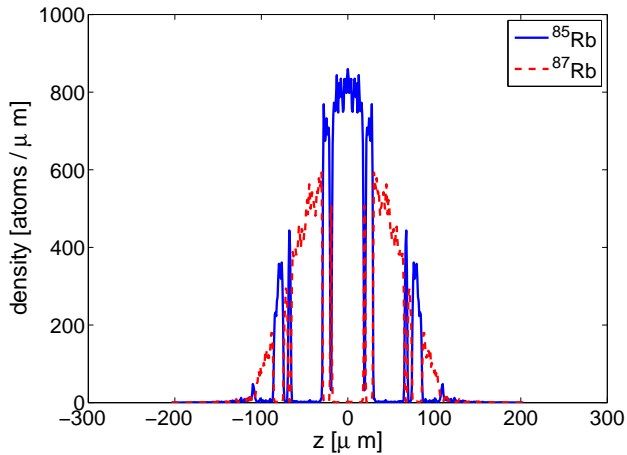}}
\caption {The final spatial pattern resulting from the dynamical
growth of the dual condensate, with scattering parameters $a_{85}=81$
bohr, $a_{87}=99$ bohr and $a_{85-87}=214$ bohr; to be compared with
Fig.~(\ref{fig:pattern3}). All other parameters are the same as in
Fig.~(\ref{fig:pattern3}).
\label{fig:pattern1}}
\end{figure}

We explored some of the factors that affect the final spatial pattern
upon the end of evaporative cooling.  As noted above, the condition
for phase separation in the Thomas-Fermi limit is $\Delta<0$ where
$\Delta$ is defined in Eq.~(\ref{eq:phase}). We find that having a
negative $\Delta$ with larger absolute magnitude leads to the
dynamical formation of \textit{fewer} bubbles.  For example,
Fig.~(\ref{fig:pattern1}) shows the modulation pattern after 1000ms
with the only difference from Fig.~(\ref{fig:pattern3}) that
$a_{85}$ has been reduced from 200 to 81 bohr, so that $\Delta$, which
for the previous case equals $-0.57$, is now $-0.82$. The number of
separated bubbles is clearly reduced, from 25 in the first case to 17
in the second. The physical reason for this effect is as follows. When
$\Delta$ is more negative (larger absolute value), the dynamical phase
separation begins earlier in the growth process when the size of the
condensates is smaller. Since the size of the condensates is smaller,
they divide into smaller number of bubbles. Saturation (that is,
complete spatial separation of the two components) is also achieved
earlier, and no further sub-division occurs at later times. On the
other hand, with smaller absolute value of $\Delta$ the condensates
have a chance to increase there size before the onset of
phase separation. Thus, eventually, when phase separation takes
place, a larger number of bubbles can be formed.

\begin{figure}
\resizebox{3.5in}{!}{\includegraphics{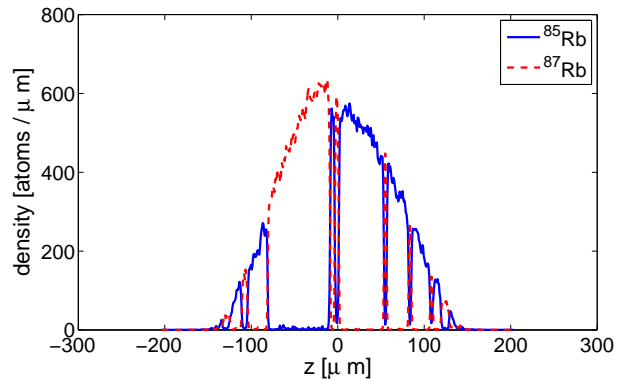}}
\caption {The final spatial pattern resulting from the dynamical
growth of the dual condensate, with parameters identical to that of
Fig.~(\ref{fig:pattern3}), except that there is a shift of 1.7 $\mu$m
between the centers of the traps of the two components.
\label{fig:pattern6}}
\end{figure}

\begin{figure*}
\resizebox{7.0in}{!}{\includegraphics{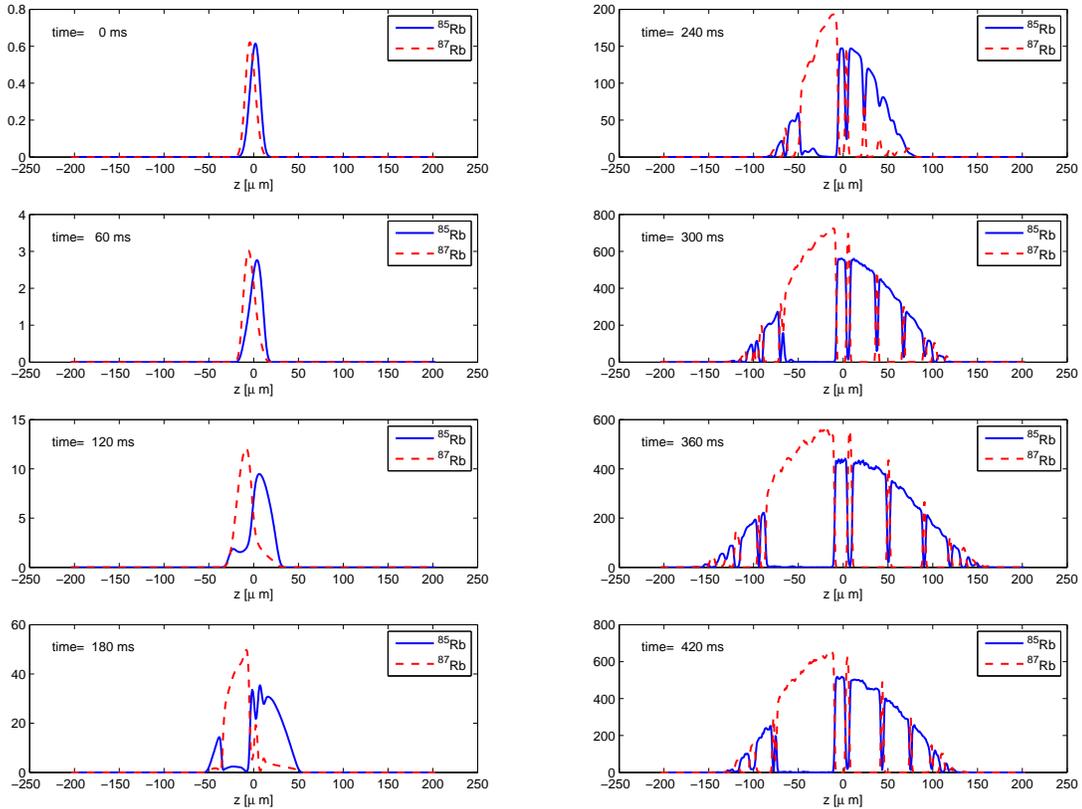}}
\caption {Dynamical growth of a dual BEC leading to the pattern
formation seen in Fig.~(\ref{fig:pattern6}).
\label{fig:dyn6} }
\end{figure*}

An even more dramatic effect is seen when the centers of the traps of
the two components are shifted one from the other along their long axis
by a small distance, as may occur due to residual gravitational sag. This can occur when the trap is slightly tilted off
the horizontal to the earth, as apparently occurred in the experiment \cite{Papp08}. Even if the tilt is slight,
the effect can be large in the axial direction due to the low and  different trapping
frequencies for the two components in this direction. This leads to asymmetry in the spatial pattern
formation. Fig.~(\ref{fig:pattern6}) shows an asymmetrical pattern
obtained with a dual condensate with the same interaction parameters
as of Fig.~(\ref{fig:pattern3}), but with the centers of the traps of
the two components shifted 1.7 $\mu$m from each
other. Fig.~(\ref{fig:dyn6}) shows the time dynamics leading to this
pattern. The initial shift is clearly seen in the partial overlap
between the ``seed'' Gaussian of the two components. As the density
grows, the inter-component repulsion causes the centers of mass of the
two components to move farther apart. At the same time, modulation
instability and bubble formation occurs where the two components
overlap. Due to the reduced spatial overlap (compared to the case with
no shift of the traps' centers), the final number of bubbles is much
reduced.

We also note that the modulation instability mechanism which is at
the heart of the spatial pattern formation requires a quasi one
dimensional geometry. In a fully 3D trap (with aspect ratio close to
one), even if ``bubbles'' are seeded, they would quickly merge and
coalesce. This is because a spatial pattern which contains many
bubbles would have higher energy then one with continuous, bulk
configuration with no bubbles, due to the contribution of surface
tension energy \cite{Schaeybroeck08}. But, in the quasi 1D trap there
is a high energy barrier for the merging of two bubbles of one
component separated by a bubble of another component. This is
essentially a tunneling process which is highly suppressed. Our
numerical simulations support this picture. For example, we find that in a trap
with aspect ratio $\lambda=\omega_r/\omega_z=10$ (compared with
$\lambda=45$ in the simulations discussed above) it is quite hard to
see bubbles, and at most four to five are formed.

Another factor in the growth dynamics concerns the start of
condensation of the two components. When the atom numbers of the two
components are the same, they have the same critical temperature and so
may be expected to begin condensation at the same time. However when
the two components contain different number of atoms, the critical
temperature of the larger component is higher ($T_c \propto
N^{1/3}$), so it may be expected to begin condensation first. The time
delay between the start of condensation of the two components will
then depend on their atom number ratio and on the rate of evaporative
cooling. We have performed simulations with large time delay such that
one component begins to condense after the end of condensation of the
second. We find that in this case the growth of the component which
condenses later occurs at the edges of the first one, and they form a
structure similar to the usual ground state, with no bubble
pattern. This may be expected since the first component to condense
creates a repulsive potential which repels the second component from
condensing in the center of the trap. Thus, a sufficient temporal overlap between
the condensation of the two components seems essential for formation of a bubble pattern.

\subsection{Comparison with experiment}

\begin{figure}
\resizebox{3.5in}{!}{\includegraphics{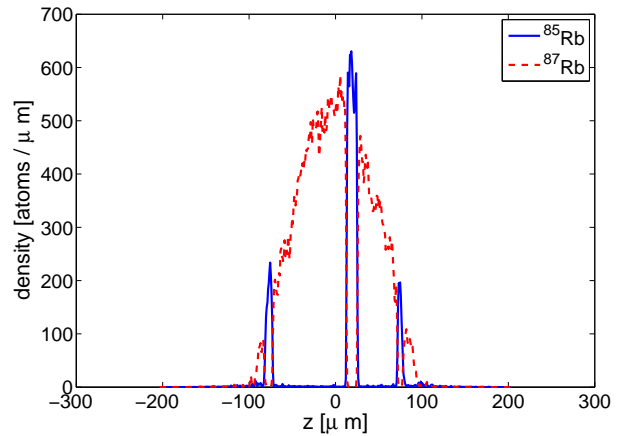}}
\caption {The final spatial pattern resulting from the dynamical
growth of a condensate with parameters approximating the experimental
conditions of \cite{Papp08}, Fig.4b there: $a_{85}=81$ bohr,
$a_{87}=99$ bohr, $a_{85-87}=214$ bohr. Trap frequencies are as in
Fig.~(\ref{fig:ground}) above.  The number of $^{85}$Rb atoms is
$10^4$ and that of $^{87}$Rb, $6 \times 10^4$. The trap centers of the
two components are shifted by $3.4$ $\mu$m from each other along the long axis of the trap, to
simulate residual gravitational sag (see text).
\label{fig:simexp} }
\end{figure}

To make a concrete connection with the experimental results
\cite{Papp08}, we simulated as close as possible the conditions under
which the remarkable bubble patterns seen in Fig.(4b) and (4c) of
Ref. \cite{Papp08} were observed.  The result is shown in
Fig.~(\ref{fig:simexp}). The number of atoms (reported in the caption)
was also estimated from the experiment \footnote{J. Pino, private
communication}, with $6 \times 10^4$ $^{87}$Rb atoms and $1 \times
10^4$ $^{85}$Rb atoms. In this simulation we assume that the two
components begin to condense together. Due to the different atom
numbers this might not have been the case in the experiment, but
should be good enough as long as the time delay between condensation
of the two components is not too long. In the simulation We can see the formation of three bubbles of $^{85}$Rb
immersed in and between the larger cloud of $^{87}$Rb. This
result is qualitatively similar to that observed in the experiment.
The comparison with the experiment is quite encouraging, especially
bearing in mind that we do not know parameters such as the exact
time(s) of the growth of the two condensates, and the initial
condensate ``seed'' sizes.  The condensates in the experiment also
show evidence for gravitational sag. In the radial direction,
gravitational sag breaks the cylindrical symmetry, an effect we have
not attempted to simulate here. However we do include the effect of
asymmetry in the axial direction due to residual gravity sag along
this direction, which can occur due a slight tilting of the trap from
the horizontal to the earth.

\section{Conclusions \label{sec:conclusions}}

In this work we have put forward a mechanism which explains the
intriguing experimental observations of \cite{Papp08}. These observations showed
the formation of a spatial modulated pattern of a dual BEC condensate
\textit{upon the end of evaporative cooling} in an highly elongated cigar
trap. The mechanism we suggest includes two crucial elements. One, the
growth of the dual condensates which is modeled by imaginary gain
terms in the Gross-Pitaevskii gain equation, Eq.~(\ref{GPGE}).
Second, as the condensates grow and their mutual interaction becomes
more and more dominant, the mechanism of modulation instability kicks
in, and gives rise to the evolution of spatial modulation with smaller
and smaller wavelength. The initial linear instabilities are amplified
by the non-linearity in the system, and this process comes to an end
when full spatial separation between the two components is
achieved. The final pattern is very long lived due to the large energy
barrier for the separated component ``bubbles'' to cross over each
other. The quasi 1D nature of the trap is essential, otherwise no
barriers exist to prevent any formed ``bubbles'' from merging and
coalescing into continuous, bulk components.

Naturally, the gain model is a simplified, phenomenological model of
the growth of two coherent condensates from the thermal gases, which
does not include the interactions between the condensates and the
thermal gases, nor any direct temperature dependence.  The thermal gas
only enters through its role as a particle reservoir (for the gain
terms). Thus, in principle, other mechanisms such as thermal
fluctuations may also play a role in the final observed spatial
pattern in the experiments. It would be of interest to have a more
elaborate (but therefore, more complicated) modeling of the
evaporative cooling process of a dual condensate. One such possible
model may be the stochastic Gross-Pitaevskii equation
\cite{Gardiner03} which recently was shown to give excellent agreement
with the experimental observation of spontaneous vortex formation in
evaporative cooling of a single component BEC \cite{Weiler08}.
Nevertheless, the encouraging agreement between our simulations and
the experiment supports our belief that the simple model put forward
here already captures the essential physics at the root of the
observed phenomenon.

\begin{acknowledgments}
S. R. and J. L. B. acknowledge financial support from NSF. M.E. and L.E.H.
acknowledge support on NSF grants PHY-0354969 and PHY-0653359.
\end{acknowledgments}

\bibliography{biblo}

\end{document}